\documentclass{article}[9pt]
\usepackage{spconf,amsmath,amssymb,graphicx}
\usepackage{cite}
\usepackage{mathrsfs}
\usepackage[lined]{algorithm2e}
\usepackage{wasysym}
\usepackage{color}
\usepackage{subfigure}
\usepackage{multirow}
\usepackage{amsthm}
\usepackage{amsopn}
\usepackage{url}
\usepackage{helvet}
\usepackage{bm}
\usepackage{rotating}
\usepackage[utf8]{inputenc}

\newcommand{\vx}{\mathbf{x}}

\newcommand{\vF}{\mathbf{F}}
\newcommand{\vW}{\mathbf{W}}

\newcommand{\vZ}{\mathbf{Z}}
\newcommand{\vX}{\mathbf{X}}
\newcommand{\vY}{\mathbf{Y}}

\title{Vocal melody extraction using patch-based CNN}
\name{Li Su}
\address{Institute of Information Science, Academia Sinica, Taiwan}
\begin{document}
\maketitle

\begin{abstract}
A patch-based convolutional neural network (CNN) model presented in this paper for vocal melody extraction in polyphonic music is inspired from object detection in image processing. The input of the model is a novel time-frequency representation which enhances the pitch contours and suppresses the harmonic components of a signal. This succinct data representation and the patch-based CNN model enable an efficient training process with limited labeled data. Experiments on various datasets show excellent speed and competitive accuracy comparing to other deep learning approaches.
\end{abstract}

\begin{keywords}
Melody extraction, convolutional neural networks, cepstrum, music information retrieval.
\end{keywords}

\section{Introduction}

Most of the music information retrieval (MIR) tasks, such as auto-tagging \cite{liu2016event} and vocal melody extraction from polyphonic music \cite{kum2016melody,rigaud2016singing,BittnerDeepSalience17,Verma2016FrequencyEF,bittner2015melody,salamon2012melody}, require local behavior of a specific audio event to be detected, e.g., the time stamps and pitches of a singing voice. The most common data representation used for such MIR tasks is undoubtedly the spectrogram obtained from converting a 1-D signal into a 2-D graph in the time-frequency domain to better capture local behaviors of a signal. However, a spectrogram is rarely modeled in a way that a conventional image processing method does: an audio processing method typically employs a frame or a segment (i.e. a sequence of frames) as a basic unit, while the counterpart for an image processing method usually being a patch, such as the 
region-based convolutional neural network (R-CNN) and its improved versions \cite{girshick2014rich, ren2015faster, he2017mask}, all of which emphasize the strategy of selecting and processing patches to bridge the gap between image classification and object localization \cite{girshick2014rich}. 

It is therefore intriguing to unlock the potential of utilizing time-frequency patches in MIR tasks considering advantages offered by employing the patches. First, a patch can capture an event localized not only in time but also in frequency or pitch. Besides, patch-based processing is faster than segment-based processing, since a patch excludes some elements in a spectrum from computation. However, modeling local pitched events on a spectrogram representation by using patches has difficulties in practice, since a pitched signal naturally has a wide-band spectrum having its harmonic series, and multiple events usually have their spectra overlapped with each other. To enjoy the advantages of patch-based modeling, one needs a data representation that can effectively localize a pitch event.

To address the above-mentioned issues, a novel data representation, based on the combined frequency and periodicity (CFP) approach \cite{su2015combining, su2016exploiting, su2017HSP_DNN, wu2018vocal}, is proposed to enable localization of pitched events in the frequency domain without interference from the harmonics. Such a data representation then allows us to leverage the key concepts in image processing, such as the region proposal and selective search in R-CNN, in MIR. This study specifically focuses on vocal melody extraction from poly-phonic music, viewing it as an analogy of semantic segmentation in image processing. The vocal melody extraction contains a classification task for classifying whether a time-frequency patch contains a vocal event, and a localization task for simultaneously performing vocal activity detection (VAD) and pitch detection. Since a singing voice has vibrato and sliding behaviors different from other instruments \cite{beauchamp2017comparison}, it is feasible to perform vocal-nonvocal classification simply by using a localized pitch contour.

The proposed system uses a similar strategy to R-CNN. It extracts a CFP representation from a signal, selects patches as candidates of vocal melody objects in the representation, trains a CNN to determine whether a patch corresponds to a singing voice or not, and then localizes a voice melody object both in time and frequency. In the experiments, the proposed system achieves competitive accuracy with respect to other recently-developed deep learning approaches by using small-sized training data and limited computing resources. 




\section{Method}
\begin{figure}[t]
\includegraphics[width=0.5\textwidth]{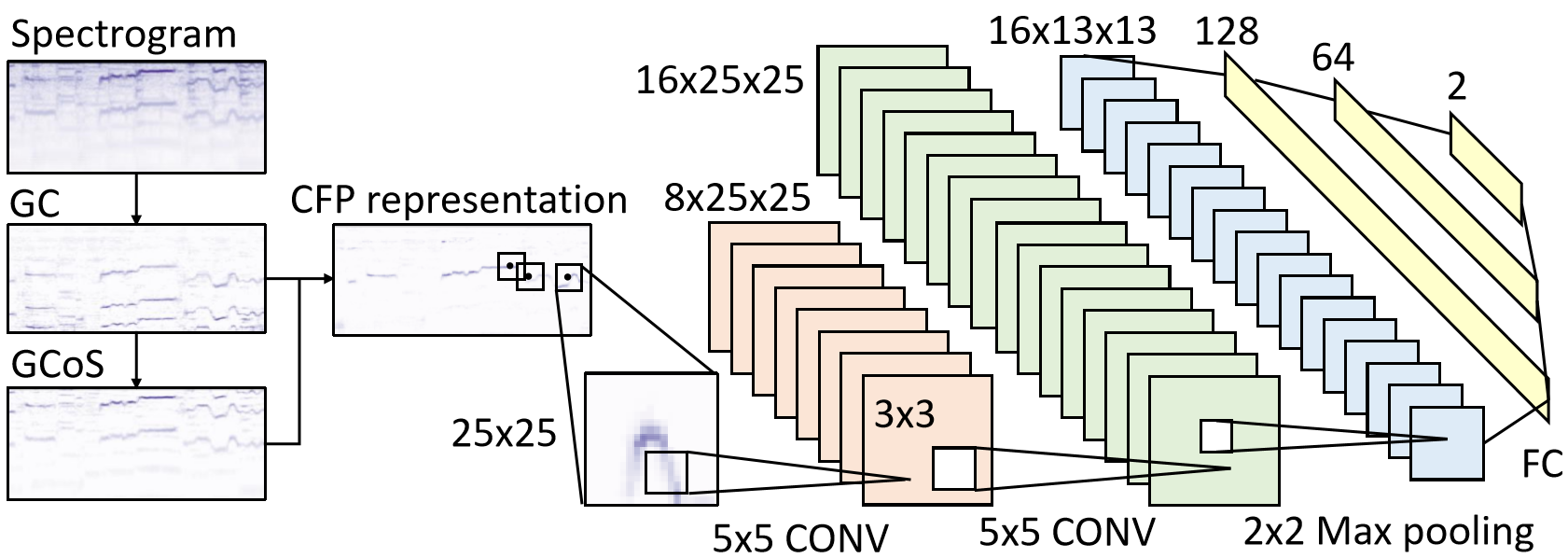}
\centering
\caption{The proposed system.}
\label{fig:system}
\end{figure}

As shown in Fig. \ref{fig:system}, the proposed melody extraction system contains three main parts: a data representation, patch selection, and CNN modeling.



\subsection{Data representation}

The data representation proposed here is based on the CFP approach, in which a pitch object is determined by a frequency-domain representation indicating its fundamental frequency ($f_0$) and harmonics ($nf_0$), and by a time-domain representation revealing its $f_0$ and sub-harmonics ($f_0/n$) \cite{su2015combining, su2016exploiting,su2017HSP_DNN,peeters2006music,wu2018vocal}. The following discussion shows that simply multiplying the frequency-domain representation by the time-domain representation can effectively suppress the harmonic and sub-harmonic peaks, and yield a CFP representation localizing a pitch object in both the time and frequency domain.

Consider an input signal $\vx:=\vx[n]$ where $n$ is the index of time. 
Let the amplitude part of the short-time Fourier transform (STFT) of $\vx$ is represented as $\vX$.
Given an $N$-point DFT matrix $\vF$, high-pass filters $\vW_f$ and $\vW_t$, and activation functions $\sigma_i$, consider the following three data representations:
\begin{align}
\vZ_0[k,n]&:=\sigma_{0}\left(\vW_f\vX\right)\,, \label{eq: specs}\\ \vZ_1[q,n]&:=\sigma_{1}\left(\vW_t\vF^{-1}\vZ_0\right)\,, \label{eq: ceps} \\ 
\vZ_2[k,n]&:=\sigma_{2}\left(\vW_f\vF\vZ_1\right)\,. \label{eq: gcos}
\end{align}
Equations (\ref{eq: specs})-(\ref{eq: gcos}) encompass many conventional pitch salience functions used in the literature: $\vZ_0$ is a spectrogram, $\vZ_1$ is a generalized cepstrum (GC) \cite{tolonen2000computationally,klapuri2008multipitch,kobayashi1984spectral,indefrey1985design,tokuda1994mel}, and $\vZ_2$ is a generalized cepstrum of spectrum (GCoS) \cite{su2017HSP_DNN,wu2018vocal}. The index $k$ in $\vZ_0$ and $\vZ_2$ is frequency, while the index $q$ in $\vZ_1$ represents {\em quefrency}, which has the same unit as time. Here the nonlinear activation function is defined as a rectified and root-power function:
\begin{equation}
\sigma_i\left(\vZ\right) = |\text{relu}(\vZ)|^{\gamma_i},\quad i=0,1,2\,,
\label{label:nonlinearfunc}
\end{equation}
where $0<\gamma_i\leq 1$, relu$(\cdot)$ represents a rectified linear unit, and $|\cdot|^{\gamma_0}$ is an element-wise root function. $\vW_f$ and $\vW_t$ are two high-pass filters designed as diagonal matrices with the cutoff frequency and quefrency, respectively being $k_c$ and $q_c$, :
\begin{equation}
\vW_{f\text{ or }t}[l,l] = \left\{
  \begin{array}{ll}
    1 \,,& \quad l>k_c\text{ or }q_c \,; \\
    0 \,,& \quad \text{otherwise}\,. \label{eq: W}
  \end{array}
  \right.
\end{equation}
$\vW_f$ and $\vW_t$ are used to remove slow-varying portions.
Based on the CFP approach, merging $\vZ_1$ and $\vZ_2$ together can suppress the unwanted harmonics and sub-harmonics. Since $\vZ_1$ is in the quefrency domain, it should be mapped into the frequency domain before merging. Besides, to fit the perceptive scale of musical pitches, the resulting CFP representation is preferred to be mapped into the log-frequency scale. Therefore, two sets of filter banks respectively in the time and frequency domains are applied. Both filter-banks have 159 triangular filters ranging from 80 Hz to 800 Hz, with 48 bands per octave. More specifically, the $m$-th filter in frequency (or time) takes the weighted sum of the components whose frequency (or period) between $0.25$ semitones above and below the frequency at $f_m=80\times 2^{(m-1)/48}$ Hz (or the period at $1/f_m$ seconds). The filtered representations are then both in the pitch scale, namely $\tilde{\vZ}_1$ and $\tilde{\vZ}_2$. The final CFP representation is  
\begin{equation}
\vY[p,n]=\tilde{\vZ}_1[p,n]\tilde{\vZ}_2[p,n]\,,
\label{eq:fusion}
\end{equation}
where $p$ is the index in the log-frequency scale. In this work, the audio files are resampled to 16 kHz and merged into one mono channel. Data representations are computed with Hann window of 2048 samples and hop size of 320 samples.

\begin{figure*}[t]
\includegraphics[width=0.9\textwidth]{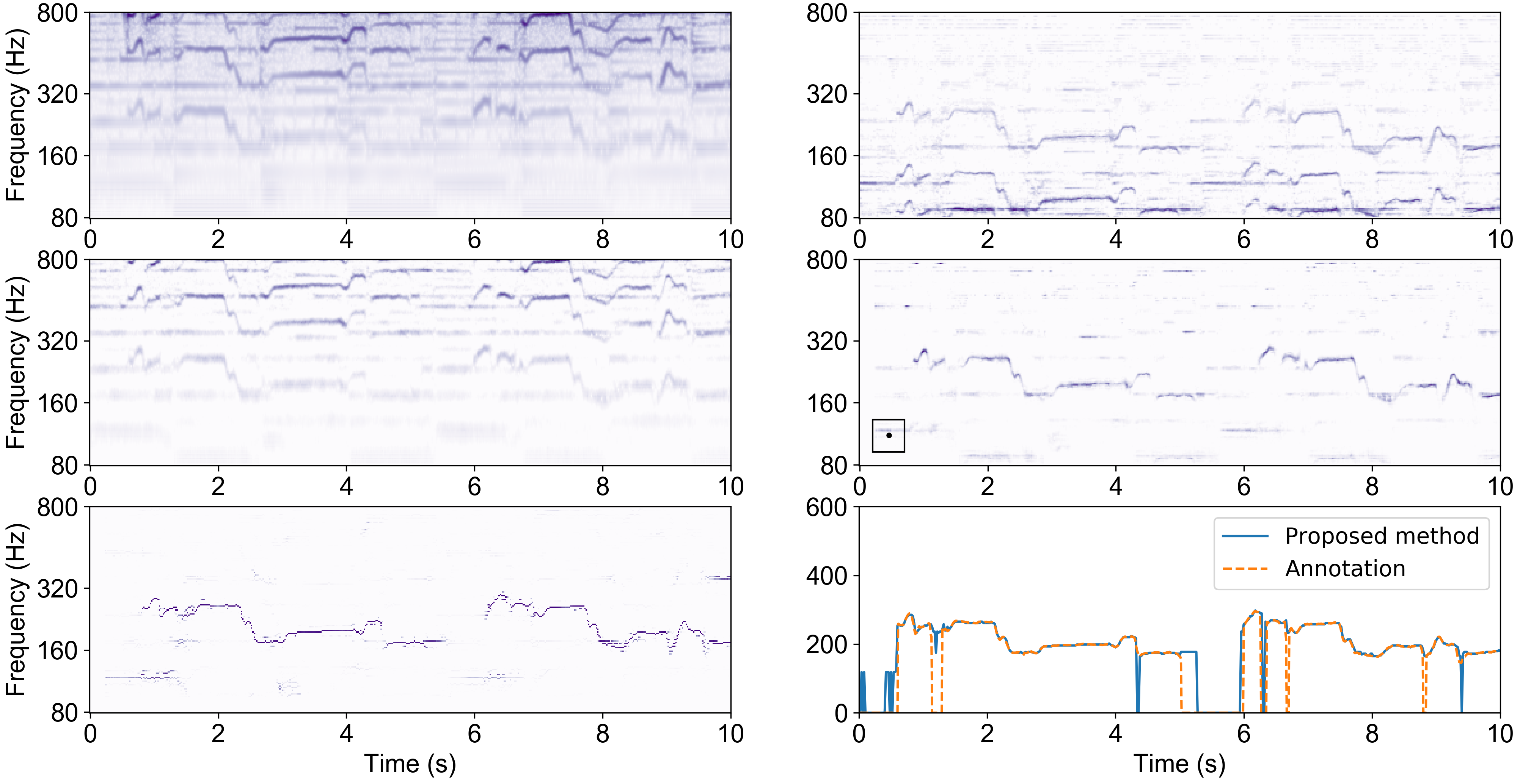}
\centering
\caption{Procedure of singing melody extraction of the first 10s of `train09.wav' in the MIREX05 dataset. Top left: power-scale spectrogram ($\vZ_0$). Top right: generalized cepstrum ($\vZ_1$). Middle left: generalized cepstrum of spectrum ($\vZ_2$). Middle right: CFP representation ($\vY$) with a $25\times 25$ patch. Bottom left: reorganized CNN outputs. Bottom right: the CNN-MaxOut result. }
\label{fig:example}
\end{figure*}

The above procedure is illustrated in the top 4 subfigures in Fig. \ref{fig:example}. As seen, the singing voice melody of the male singer performing in this clip has very weak and unclear con-tour at its fundamental frequency, and the spectral energy is mostly concentrated in the high frequency range. Such a ‘weak fundamental’ phenomenon is usually seen in singing voice because the energy of singing voice is dominated by the formants rather than the fundamental frequency. Such high-frequency contents are unseen in the GC representation as it is a time-domain representation, and the GC representation exhibits a number of strong sub-harmonics in the low frequency range. The GCoS is then a refined, ‘de-trended’ version of the spectrogram, where the spectral envelope is canceled by the high-pass filter in (\ref{eq: W}). Multiplying GCoS by GC therefore yields a succinct CFP representation in (\ref{eq:fusion}), in which the unwanted peaks are suppressed, and the remaining peaks are mostly the fundamental frequencies of either the melody or the accompaniment instruments. 

\subsection{Patch selection}

To classify whether a pixel in a time-frequency plane belongs to a singing voice melody contour, patches from the key regions are selected in $\vY$ as the candidates of a melody contour. Assuming that every peak in a frame in $\vY$ is a candidate of vocal melody, a simple strategy is employed for selecting a patch: for each peak, a patch having the peak at its center is selected. The patch size is $25\times 25$ (i.e., 0.5 second 6.25 semitones) as suggested by our pilot study, and a fixed size is employed for simplicity. The patches are then used as input of a CNN trained to classify whether the center of a patch belongs to a vocal melody contour (labeled as 1) or not (labeled as 0). Only a small portion of the peaks in $\vY$ are the melody contour, and therefore 10 percent of non-vocal peaks are randomly selected into the training data during training to avoid the imbalance between positive and negative examples. 

\subsection{Model}

To predict whether an input patch represents singing voice or not, we configure a CNN model having two convolutional layers followed by three fully-connected layers. The two convolutional layers have 8 $5\times 5$ filters and 16 $3\times 3$ filters, respectively, and both preserve the size of the mapping with padding. The numbers of units of the three fully-connected layers are 128, 64, and 2, respectively. The CNN outputs a $2\times 1$ vector with one element representing the likelihood of being a melody contour, and the other element representing of the opposite. During training, the cross entropy between the CNN output and the ground truth is minimized using stochastic gradient descent with the Adam optimizer. The CNN model is implemented in Python 2.7, using Keras with Theano as the back end.

The patch-level prediction results are then reorganized as a new time-frequency representation according to the time and frequency of the patch center, as shown in the bottom left of Fig. \ref{fig:example}. Note that pitch contours having low similarity to those of singing voice, such as a pitch contour having no vibrato or sliding, are of very low output probability. To generate a binary prediction result, the threshold of the CNN output is set at 0.5. As there might be more than one patch having an output probability $>0.5$, the output vocal melody contour for each time step can be obtained in three possible ways: 1) directly employing the CFP representation by simply taking the pitch index corresponding to the maxima of the frame, 2) from patches having an output probability $>0.5$, taking the frequency index where the CFP representation reaching maximal, and 3) taking the frequency index corresponding to the largest output probability. In the following discussions, we name the three methods as CFP-Max, CNN-MaxIn, CNN-MaxOut, respectively.

\section{Experiment}

We compare the three proposed methods with two recently-developed deep learning algorithms The first is the multi-column DNN (MCDNN) by Kum {\em et al.}, in which the results of ADC2004 and MIREX2005 are reported \cite{kum2016melody}, and the second is the deep salience map (DSM) by Bittner {\em et al.}, for which on-line source code with the `vocal' option is available \cite{BittnerDeepSalience17}. Since the detection results of DSM are sensitive to the thresholding parameter, the parameter is tuned from 0 to 0.9 for all datasets to find the optimal value for better comparison. The resulting optimal threshold th=0.1 as well as the default value th=0.3 are both used in the experiment. The model is trained and tested on an ASUS ZenBook UX430U @ 2.70/2.90 GHz with an Intel i7-7500U CPU and 8 GB RAM. For reproducibility, the companion source code, and information of testing data and results can be found at \url{https://github.com/leo-so/VocalMelodyExtPatchCNN}.

\begin{table*}[t]
\small
\centering
\label{tab:result}
\begin{tabular}{|l|ccc|cc|ccc|cc|cc|cc|}
\hline
\multirow{2}{*}{Method} & \multicolumn{5}{c|}{ADC2004 (vocal)} & \multicolumn{5}{c|}{MIREX2005 (vocal)} & \multicolumn{2}{c|}{iKala} & \multicolumn{2}{c|}{MedleyDB} \\
\cline{2-15}
 & OA & RPA & RCA & VR & VFA & OA & RPA & RCA & VR & VFA & RPA & RCA & RPA & RCA\\
\hline
CFP-Max & 61.2 & 71.7 & 76.8 & 100 & 100 & 46.3 & 70.7 & 75.5 & 100 & 100 & 69.7 & 72.6 & 55.6 & 62.4 \\
CNN-MaxIn  & \textbf{74.3} & 76.7 & 78.4 & 90.1 & 41.3 & 73.2 & 81.2 & 82.2 & 95.1 & 41.1 & 76.6 & 77.8 & 58.7 & 63.6 \\
CNN-MaxOut & 72.4 & 74.7 & 75.7 & 90.1 & 41.3 & 74.4	& \textbf{83.1} & \textbf{83.5} & 95.1 & 41.1 & \textbf{76.9} & \textbf{77.7} & 59.7 & 63.8 \\ \hline
MCDNN     & 73.1 & 75.8 & 78.3 & 88.9 & 41.2 & 68.4 & 77.6 & 78.6 & 87.0 & 49.0 & -- & -- & -- & -- \\
DSM (th = 0.3) & 68.0 & 68.4 & 70.9 & 78.2 & 25.5 & \textbf{76.3} & 70.4 & 71.2 & 80.1 & 13.6 & 67.9 & 69.7 & 61.7 & 64.7 \\
DSM (th = 0.1) & 70.8  & \textbf{77.1} & \textbf{78.8} & 92.9 & 50.5 & 69.6 & 76.3 & 77.3 &   93.6 & 42.8 & 73.4 & 74.6 & \textbf{72.0} & \textbf{74.8} \\
\hline
\end{tabular}
\caption{Vocal melody extraction results of the proposed and other methods on various datasets.}
\end{table*}



\subsection{Data}

The first 3 seconds of the first 800 clips of the MIR1K\footnote{https://sites.google.com/site/unvoicedsoundseparation/mir-1k} dataset are used as training data, and no data augmentation is performed. In contrast to other data-driven melody methods, the training data used for this study is small-sized and of low-diversity, since it contains only 40 minutes of Chinese karaoke songs sung by amateur singers. 
The testing data for evaluation are from four datasets for melody extraction: ADC2004, MIREX05,\footnote{https://labrosa.ee.columbia.edu/projects/melody/} iKala \cite{chan2015vocal}, and MedleyDB \cite{bittner2014medleydb}. As the proposed model is designed solely for singing voice melody, we follow \cite{kum2016melody} and select only samples having melody sung by human voice from ADC2004 and MIREX05. As a result, 12 clips in ADC2004 and 9 clips in MIREX05 are selected. Since all melodies in iKala are singing voice, all 252 clips in the iKala dataset are selected for evaluation. To obtain the annotation of singing voice in medleyDB, 12 songs having singing voice included in their `MELODY2' annotations are selected. The vocal melody labels are obtained from the MELODY2 annotations occurring in the intervals labeled by `female singer' or `male singer'. 

\subsection{Result}

Table \ref{tab:result} lists the overall accuracy (OA), raw pitch accuracy (RPA), raw chroma accuracy (RCA), voice recall (VR) and voice false alarm (VFA) of the three proposed methods, MCDNN, and DSM on the four testing datasets. All data is computed from mir\_eval \cite{raffel2014mir_eval}. The results of MCDNN on ADC2004 and MIREX05 are from  \cite{kum2016melody}. Due to paper length limitation, we only report the VR and VFA of the ADC2004 and MIREX05 datasets since the VAD task is a minor topic of this paper. 
CFP-Max has no VAD mechanism, so the resulting VR and VFA are both 100\%. CNN-MaxIn and CNN-MaxOut both use the same CNN for VAD, so their VR and VFA values are the same. 
The VR and VFA of the proposed model are similar to the ones of the DSM with the optimal threshold parameter without fine tuning; it is probably because each instance of voice activation is determined independently and locally in both the time and frequency domain.

Among the proposed methods, CNN-MaxOut achieves highest scores in general; this result clearly verifies the effectiveness of the proposed patch-based CNN modeling in selecting the patches corresponding to singing voice melody. However, one can also observe that the CFP representation itself, as a pure signal processing method without data-driven modeling, still exhibits very competitive performance. The explanation for this phenomenon includes: first, using CFP-Max without CNN modeling provides very competitive RPA and RCA, occasionally even better than DSM using the default threshold; second, the RPA and RCA of CNN-MaxIn are higher than those of CNN-MaxOut in the ADC2004 dataset.


For the comparison to other methods, Table \ref{tab:result} shows that when focusing on RPA, CNN-MaxOut outperforms DSM by 6.8\% in MIREX05 and by 3.5\% in iKala. On the other hand, DSM outperforms CNN-MaxIn by 0.4\% in ADC2004 and also outperforms CNN-MaxOut by 12.3\% in MedleyDB. Such a huge difference across the four datasets encompassing a broad range of music styles can be attributed to the training method and the size of the model. DSM is trained with data most similar to MedleyDB, so its superior performance is no surprise. Similarly, ADC2004 contains opera singing included in the training data of MedleyDB but not included in MIR-1K, so for those samples of opera singing, DSM also performs better than the model proposed in this study. On the other hand, although the proposed model is trained on amateur singing voice, it still performs better in iKala, a dataset with professional Karaoke singing. For the size of model, DSM adopts more layers, more feature mappings, and larger data representation (i.e., the full CQT representation) than the proposed model, which has a small, patch-sized data representation and fewer feature mappings. As a result, the proposed model provides a much faster computation speed than DSM, and a test on the first 10 songs in the iKala dataset indicates the execution speeds of CFP-Max, CNN-MaxOut and DSM are 0.10x, 0.57x and 4.26x real time, respectively, demonstrating that the proposed patch-based CNN model is around eight times faster than DSM.

\section{Conclusion}
The proposed model with the CFP representation and patch-based CNN is demonstrated as easy to train, computationally light, and of high accuracy. It also has scalability, and can be extended in the future by incorporating data augmentation and more training samples containing melody contours locally representing voice or non-voice events.

The solution proposed is based on the assumption that a short pitch contour is sufficient to discriminate singing voice melody from accompaniment; this might no longer hold under complicated instrumentation or other context-dependent situation. Future research directions can include incorporating both the local and contextual information spanning either in time or frequency to address more challenging scenarios.


\bibliographystyle{IEEEbib}
\bibliography{vocal_ext}

\end{document}